\documentclass[showpacs,pra,twocolumn]{revtex4}
\usepackage{hyperref}
\usepackage{amssymb}
\usepackage{amsmath}
\usepackage{graphicx}

\begin{document}
% Header 
\title{Quantum sensitivity limit of a Sagnac  
hybrid interferometer based on slow-light propagation in ultra-cold gases}
\author{F.~E.~Zimmer and M. Fleischhauer}
\affiliation{Fachbereich Physik der Technischen
 Universit{\"a}t Kaiserslautern, D-67663\\
Kaiserslautern, Germany}
\date{\today}
%%%%%%%%%%%%%%%%%%%%%%%%%%%%%%%%%%%%%%%%%%%%%%%%%%%%%%%%%%%
%
% Abstract
%
%%%%%%%%%%%%%%%%%%%%%%%%%%%%%%%%%%%%%%%%%%%%%%%%%%%%%%%%%%%
\begin{abstract}
The light--matter-wave Sagnac interferometer based on ultra-slow light proposed recently in (Phys.~Rev.~Lett. \textbf{92}, 253201 (2004)) is analyzed in detail. 
In particular the effect of confining 
potentials is examined and it is shown that the ultra-slow light attains a rotational
phase shift equivalent to that of a matter wave, if and only if 
the coherence transfer from light to atoms associated with slow light is associated with
a momentum transfer and if an ultra-cold gas in a ring trap is used. The quantum sensitivity limit of the
Sagnac interferometer is determined and the minimum detectable rotation rate 
calculated. It is shown that the slow-light interferometer allows for a significantly higher 
signal-to-noise ratio as possible in current matter-wave gyroscopes.
\end{abstract}
\maketitle
%%%%%%%%%%%%%%%%%%%%%%%%%%%%%%%%%%%%%%%%%%%%%%%%%%%%%%%%%%%
%
% ARTICLE
%
\section{Introduction}
%
%%%%%%%%%%%%%%%%%%%%%%%%%%%%%%%%%%%%%%%%%%%%%%%%%%%%%%%%%%%
In contrast to inertial motion, rotation of an
object is absolute in the sense that it can be defined intrinsically, i.e.
independent of any inertial frame of reference. Rotation
can be detected e.g. by means of the Sagnac effect
 \cite{Sagnac-CRAS-1913}, i.e. the relative phase shift $\Delta\phi_{\rm rot}$ of 
counterpropagating
waves in a ring interferometer of area $\mathbf{A}$ 
attached to the laboratory frame rotating with angular velocity 
$\mathbf{\Omega}$. 
\begin{equation}
\Delta\phi_{\rm rot} = \frac{4\pi}{\lambda v} \mathbf{\Omega}\cdot\mathbf{A},
\end{equation}
where $\lambda$ is the wavelength and $v$ the phase velocity of the 
wave.
Depending on the nature of
the wave phenomena employed, one distinguishes two basic types of
Sagnac interferometer: laser \cite{Post-RMP-1967,Chow-RMP-1985,Stedman-RPP-1997} and matter-wave gyroscopes \cite{Bongs-RPP-2004}. 
The Sagnac phase shift per unit area in a 
matter-wave device exceeds that of laser-based gyroscopes 
by the ratio of rest energy per particle to photon 
energy $mc^2/\hbar\omega$ which for alkali
atoms and optical photons is of the order of $10^{11}$
\cite{Scully-1997,Page-PRL-1975}. 
Despite this very large number, matter-wave gyroscopes have only recently
reached the short-time sensitivities of laser based devices 
\cite{Gustavson-PRL-1997,McGuirk-PRL-2000}. This has mainly two reasons:
First of all, laser-based gyroscopes, especially fiber-optics interferometer, can 
have a much larger area than matter-wave systems \cite{Culshaw-MST-2006}. Secondly
the large flux of photons achievable in optical systems leads to a much lower 
shot-noise level as compared to matter-wave devices \cite{Chow-RMP-1985,Kasevich-Science-2002}.
Thus in order to make full use of the much larger rotational sensitivity
per unit area in a matter-wave device one needs to find ways to 
increase {\it (i)} the interferometer area and {\it (ii)} the particle flux.
While a substantial increase of the interferometer area 
in matter-wave devices is difficult, the use of
novel cooling techniques has lead to high-flux atom sources which improved the performance of atom interferometers
\cite{Orzel-Science-2001,Bongs-RPP-2004}. With particle throughputs which can now reach 
$10^8$ s$^{-1}$ as compared to a few atoms per second %s$^{-1}$ 
in the first atomic
interferometers, the noise level is however still
much higher than that achievable in fiber optics gyroscopes 
with photon counting rates on the order of $10^{16}$ s$^{-1}$ 
\cite{Kasevich-Science-2002,Bongs-RPP-2004}. 
Continuously loaded Bose-Einstein condensates
(BEC) could provide a source for coherent atoms with larger
flux values, and
substantial progress has been made in this direction over the past 
few years \cite{Chikkatur-Science-2002}.

We recently proposed a light--matter-wave hybrid interferometer based on
slow-light propagation in ultra-cold gases of
3-level atoms \cite{Zimmer-PRL-2004}. We argued that this interferometer
would combine the large rotational phase shift of matter-wave systems
with the large area typical for optical gyroscopes. To this end
the simultaneous 
coherence and momentum transfer associated with ultra-slow light in 
cold atomic gases with electromagnetically induced transparency (EIT)
\cite{Fleischhauer-RMP-2005} was utilized. As the reduction of the group velocity
of light in 3-level EIT media is based on the change of character of the dressed 
eigenmodes of the systems from electromagnetic excitations to atomic Raman excitations 
\cite{Fleischhauer-PRL-2000}, light waves can coherently be transformed
into matter waves. These matter waves pick up a Sagnac phase shift per unit area
which is orders of magnitude larger than the corresponding value for electromagnetic fields.

In the present paper we  present a detailed
theoretical description of the light-matter-wave hybrid interferometer. In particular we
discuss the effect of confining potentials for the atoms. We find that 
in contrast to the case of an infinitely extended medium or of periodic boundary conditions, which have been
assumed in \cite{Zimmer-PRL-2004}, the wavefunctions of all
three internal states acquire the same matter-wave contribution
to the Sagnac phase when in motional equilibrium with a trapping potential
\cite{Hendriks-QOPT-1990}. 
As a consequence the
matter-wave contribution to the rotational phase shift vanishes. Only if periodic 
boundary conditions for the ground-state
wavefunction can be maintained a nonvanishing
matter-wave contribution to the rotational phase shift emerges. This can be
realized e.g. in a circular-waveguide 
BEC \cite{Gupta-PRL-2005,Arnold-PRA-2006}. 
% Alternatively or if sufficiently fast changes in the 
% rotation rate are considered i.e. in the case of a sensor for rotational 
% acceleration. 
The need for a circular atomic waveguide puts more stringent 
restrictions
to the possible interferometer area then assumed in \cite{Zimmer-PRL-2004}
and thus partially invalidates the advantages of the hybrid interferometer stated
in that paper. We will show however, that despite this restriction the minimum
detectable rotation rate at the shot-noise limit can exceed the current state of the art. 
It corresponds to that of a 
matter-wave gyroscope with a rather large particle flux 
given by  the high density of 
the ultra-cold gas, e.g. a BEC, multiplied by the recoil velocity. 
To determine the quantum sensitivity limit of the hybrid interferometer
the saturation of the Sagnac phase shift with the probe-light
intensity as well as  probe-field absorption will be taken into account.
It will be shown that the Sagnac phase attains a maximum value for
a certain value of the probe power. Optimum parameter values for a maximum
signal-to-noise ratio (SNR) will be determined and the minimum detectable
rotation rate $\Omega_{\rm min}$ per unit area derived. 
%%%%%%%%%%%%%%%%%%%%%%%%%%%%%%%%%%%%%%%%%%%%%%%%%%%%%%%%%%%
%
\section{Dynamics in the rotating frame}
%
%%%%%%%%%%%%%%%%%%%%%%%%%%%%%%%%%%%%%%%%%%%%%%%%%%%%%%%%%%%

\begin{figure}[hbt]
\includegraphics[width=7.5cm]{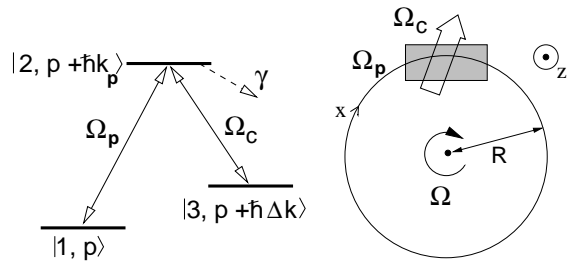}
\caption{({\it left})  atomic level scheme. 
$p$ denotes the  momentum of the atoms and $k_p$ the wavevector of the probe-field 
along the periphery ($x$) of the circular loop depicted in the right part
of the figure. $k_c^\parallel$ in the component of the control-field wavevector along
$x$, and $\Delta k= k_p-k_c^\parallel$.
({\it right}) Schematic set-up of the hybrid Sagnac interferometer with
vapor cell (grey box) attached to the rotating frame with angular velocity $
{\Omega}$.}
\label{fig:system}
\end{figure}

An intrinsic sensor attached to the laboratory frame detects the rotation of the frame
without any reference to some other, non-rotating frame of reference.
Thus it is most natural to describe this system from the point of view of a co-rotating observer.
We will give here a microscopic description of the gyroscope which consists of 
an ensemble of three-level atoms with internal states
$|1\rangle$, $|2\rangle$, and $|3\rangle$ in a ring interferometer, coupled 
by two laser
fields with (complex) Rabi frequencies $\Omega_c$ and $\Omega_p$ 
in a Raman configuration as shown in Fig.~\ref{fig:system} (\emph{left}).
The probe field $\Omega_p$ is assumed to propagate clockwise and counter-clockwise
with respect to the rotation axis $\mathbf{e}_z$ with its beam path bound to a circle of radius $R$
as depicted in Fig.~\ref{fig:system} (\emph{right}). The control field $\Omega_c$, which is assumed
to have a much larger Rabi frequency than the probe field, propagates in a different
direction such that the corresponding wavevectors are (nearly) perpendicular.
The ensemble as well as the
laser sources are attached to the laboratory frame rotating with angular
velocity $\mathbf{\Omega(t)}=\Omega(t) \mathbf{e}_z$ \cite{Dufour-CRAS-1937}.
The center-of-mass motion
of the atoms shall also be confined to the periphery of the circular 
loop.
Furthermore, it is assumed that $|\Omega| R \ll c$ such that non-relativistic quantum 
mechanics applies. 
%%%%%%%%%%%%%%%%%%%%%%%%%%%%%%%%%%%%%%%%%%%%%%%%%%%%%%%%%%%%%%%%%%%%%%%
 
%%%%%%%%%%%%%%%%%%%%%%%%%%%%%%%%%%%%%%%%%%%%%%%%%%%%%%%%%%%%%%%%%%%%%%%%%
Under conditions of two-photon resonance, the control field $\Omega_c$
generates EIT for the probe field associated with a substantial 
reduction of its group velocity \cite{Hau-Nature-1999,Budker-PRL-1999,Kash-PRL-1999}. 
The group velocity reduction which is due to
the coupling of the probe light to the atomic Raman coherence corresponds in
a quasi-particle picture to the formation of so-called dark-state polaritons,
a superposition of light and matter degrees of freedom
\cite{Fleischhauer-PRL-2000,Matsko-AdvAMO-2001}.  
The smaller the group velocity the larger the contribution of the matter component 
in the polariton. 

The atoms are here described in second quantization by three Schr\"odinger fields
$\hat\Psi_1(\textbf{r},t)$, $\hat\Psi_2(\textbf{r},t)$, and 
$\hat\Psi_3(\textbf{r},t)$ corresponding to the three internal states. 
In order to describe the propagation of the probe light  and the
three matter-wave fields in the co-rotating frame, we need to transform the
Hamiltonian of the system into the rotating frame.

As the starting point we choose the standard atom-light 
interaction Hamiltonian 
of quantum optics in Coulomb gauge and after the Power-Zienau-Wolley
 transformation 
\cite{Cohen-Tannoudji-1997a}. Adding the free Hamiltonian of a 
3-component
non-relativistic Schr\"odinger field, the Hamiltonian reads in a 
non-rotating frame 
\begin{align}
&\hat{H}=\hat{H}^{(A)}+\hat{H}^{(F)}+\hat{H}^{(I)}\nonumber\\
&\enspace =\sum_\mu\int\!{\rm d}^3 r\, 
  \hat\Psi_\mu^\dagger ({\mathbf r})
  \biggl[-\frac{\hbar^2}{2m}{\pmb\nabla}^2+
\hbar\omega_\mu+V_{\mu}^{\rm ext}({\mathbf r},t)\,\biggr]
  \hat\Psi_\mu({\mathbf r})\nonumber\\
&\enspace +\frac{\epsilon_0}{2}\int{\rm d}^3 r\,
 \left[
  \left(\frac{\hat{\mathbf\Pi}({\mathbf r})}{\epsilon_0}\right)^2
  +c^2\left({\pmb\nabla}\times\hat{\mathbf A}_{\bot}({\mathbf r})\right)^2
  \label{eq:InertialFrameSystemOperator}
 \right]\\
 &\enspace +\sum_{\mu,\nu}\int{\rm d}^3 r\, 
  \hat\Psi^\dagger_\mu ({\mathbf r})
  \left[ 
  {\mathbf d}_{\mu\nu}\cdot\left(\hat{{\mathbf\Pi}}({\mathbf r})-
{\mathbf E}_{\rm ext}({\mathbf r},t))\right)
  \right] \hat\Psi_\nu({\mathbf r}).
  \nonumber
\end{align}
$\hat{H}^{\rm (A)}$ describes the motion of atoms in an external, possibly state- and time-dependent trapping potential $V^{\rm ext}_\mu(\mathbf{r},t)$, $E_\mu=\hbar\omega_\mu$ is the energy of atoms in 
state $|\mu\rangle$. 
The free Hamiltonian of the radiation field is denoted by $\hat{H}^{\rm (F)}$,
where $\hat{\mathbf A}_{\bot}({\mathbf r})$ is the transverse part of the
vector potential and $\hat{\mathbf\Pi}({\mathbf r})$
its conjugate
momentum.
Finally $\hat{H}^{\rm (I)}$ describes the interaction  of the atoms with 
the quantized electromagnetic field and additional external fields in dipole approximation, where ${\mathbf d}_{\mu\nu}$ is the vectorial dipole matrix element between internal
states $|\mu\rangle$ and $|\nu\rangle$.
For notational simplicity we will drop 
the subscript $''\bot''$ in the following.

The transition to a frame rotating with angular velocity ${\mathbf\Omega}(t)$
is done via the unitary transformation
\begin{equation}
U(t)=\exp\left( -\frac{\rm i}{\hbar}\int_{t_0}^t\!\! {\rm d}\tau\,\,{\mathbf\Omega}(\tau)
\cdot\hat{{\mathbf L}}\right),
\label{eq:UnitaryTransToRotaFrame}
\end{equation} 
where $\hat{{\mathbf L}}$ is the total angular momentum operator 
of light and matter. By choosing ${\mathbf \Omega}=
\Omega\,{\mathbf e}_z$ we restrict ourselves to a rotation about the
fixed $z$-axis. In this case only the vector component parallel to that axis, i.~e.
\begin{align}
\hat{L}_{z}&=\hat{L}_{z}^{(\rm A)}+\hat{L}_{z}^{(\rm F)}\nonumber\\
&=\frac{\hbar}{\rm i}\sum_\mu\int\!\!{\rm d}^3r\, \hat\Psi_\mu^\dagger
\partial_{\varphi}\hat\Psi_\mu
\label{eq:TotalAngularMomentumOperatorZ}\\
&-\frac{1}{2}\sum_j\int\!\! {\rm d}^3 r\,
\left[\hat{{\Pi}}_{j}(\partial_{\varphi}
\hat{{A}}_{j})+(\partial_{\varphi}\hat{{A}}_{j})
\hat{{\Pi}}_{j}\right]\nonumber
%\rho{\rm d}\rho{\rm d}\varphi{\rm d}z.
%\dot{\hat{{\mathbf A}}}_{\bot}
\end{align}
is relevant. In eq.~(\ref{eq:TotalAngularMomentumOperatorZ})
the index $\mu$ denotes the three internal states and the index $j$
the three spatial dimensions.
The Hamiltonian in the rotating frame is hence given by 
\begin{equation}
\hat{H}_{\rm rot}=U(t)\,\hat{H}\,U^\dagger (t)+{\Omega}(t)\hat{L}_z.
\label{eq:HamilOperaInRotaFrame}
\end{equation}
Since $\hat{L}_{z}^{(\rm A)}$ and $\hat{L}_{z}^{(\rm F)}$ commute, 
the unitary 
transformation eq.~(\ref{eq:UnitaryTransToRotaFrame}) can be decomposed 
into two operators 
which act on the matter-wave and on the electromagnetic field 
respectively. One finds
\begin{align}
&\hat{H}_{\rm rot}^{\rm (A)}=\Omega(t) \hat{L}_{z}^{\rm (A)}+\\
&+
\sum_\mu\int{\rm d}^3 r'\hat\Psi^\dagger_\mu
({\mathbf r}')
\biggl[-\frac{\hbar^2}{2m}{\pmb\nabla'}^2+\hbar\omega_\mu
+V^{\rm ext}_\mu\left(
{\mathbf r}'\right)\biggr]\hat\Psi_\mu({\mathbf r}'),
\label{eq:AtomHamiOperInRotaFrame}\nonumber\\
&\hat{H}_{\rm rot}^{\rm (F)} =\Omega(t) \hat L_{z}^{\rm (F)}+\hat{H}_{0}^{(F)}\\
&+\sum_{\mu,\nu}\int{\rm d}^3 r'\, 
  \hat\Psi^\dagger_\mu ({\mathbf r'})
  \left[ 
  {\mathbf d}_{\mu\nu}\cdot\left(\hat{{\mathbf\Pi}}({\mathbf r'})-
{\mathbf E}_{\rm ext}({\mathbf r'},t)\right)
  \right] \hat\Psi_\nu({\mathbf r'}).\nonumber
\end{align}
Here the prime denotes that the variables are given with respect
 to the rotating coordinates
\begin{eqnarray}
{\mathbf r}^\prime &=& {\mathbf r}+\int_{t_0}^t\!\! {\rm d}\tau\, 
{\mathbf e}_\varphi \, R\,\Omega(\tau),
\end{eqnarray}
with $R$ being the distance from the rotation axis.
For all field operators $\hat {\mathbf F}\in 
\{\hat{\Psi},\hat{\mathbf \Pi},\hat {\mathbf A}\}$ holds:
\begin{equation}
U\hat {\mathbf F}({\mathbf r})U^\dagger=\hat{ \mathbf F}\Bigl({\mathbf r}+\int_{t_0}^t\!\! {\rm d}\tau\,\,
{\mathbf e}_{\varphi}\, R\, \Omega(\tau)\Bigr).
\label{eq:TranMattWaveAmp}
\end{equation}

The center-of-mass dynamics of the matter-wave fields is then governed by the 
following Heisenberg equations of motion in the co-rotating frame
\begin{align}
{\rm i}\hbar\Bigl(\partial_t  & + \,\, \Omega(t)\, \partial_\varphi\Bigr)
\hat\Psi_\mu ({\mathbf r}^\prime,t) = \\
&=\biggl[-\frac{\hbar^2}{2m}{\pmb\nabla}^{\prime \, 2}+\hbar\omega_\mu
+V^{\rm ext}_\mu({\mathbf r}^\prime)\biggr] \hat\Psi_\mu({\mathbf r}^\prime,t)\nonumber\\
&+\sum\limits_\nu {\mathbf d}_{\mu\nu}\cdot \left( \hat{\mathbf\Pi}({\mathbf r}^\prime)-
{\mathbf E}_{\rm ext} ({\mathbf r}^\prime,t)\right)
 \hat\Psi_\nu({\mathbf r}^\prime,t).
\label{eqn:MatWavEquMotRotFra}
\end{align}
Correspondingly the equations of motion for the conjugate momentum 
$\hat{{\mathbf\Pi}}$ and the transversal vector potential $\hat{\mathbf A}$ read
\begin{align}
\Bigl(\partial_t+\Omega(t)\partial_{\varphi^\prime}\Bigr)
\hat{{\mathbf\Pi}}({\mathbf r}^\prime,t)=
-\frac{1}{\mu_{\rm 0}}{\pmb\nabla}^\prime\times ({\pmb\nabla}^\prime
\times\hat{{\mathbf A}}({\mathbf r}^\prime,t)),
\label{eqn:MaxwellEqPotentials1}
\end{align}
and
\begin{align}
\Bigl(\partial_t+\Omega(t)\partial_{\varphi^\prime}\Bigr)
\hat{{\mathbf A}}({\mathbf r}^\prime,t)=
\frac{1}{\epsilon_{\rm 0}}\hat{{\mathbf\Pi}}({\mathbf r}^\prime,t)
+\frac{1}{\epsilon_{\rm 0}}\hat{{\mathbf P}}({\mathbf r}^\prime,t).
\label{eqn:MaxwellEqPotentials2}
\end{align}
In eq.~(\ref{eqn:MaxwellEqPotentials2}) we have introduced the transversal 
polarization $\hat{{\mathbf P}}({\mathbf r},t)=
\sum_{\mu,\nu}\hat\Psi^\dagger_\mu({\mathbf r},t)\,{\mathbf d}_{\mu\nu}
\hat\Psi_\nu({\mathbf r},t)$. 
It is immediately obvious that the transformation to the
rotating frame just amounts to the replacement $\partial_t \, \longrightarrow
\partial_t +\Omega(t)\, \partial_\varphi$.
For notational simplicity we will omit in the following the prime that indicates
rotating coordinates.

As we work in the Coulomb  gauge 
we have $\hat{{\mathbf\Pi}}({\mathbf r},t)=
-\hat{\mathbf D}({\mathbf r},t)$ \cite{Cohen-Tannoudji-1997a}.
 Using this and $\hat{{\mathbf D}}({\mathbf r})
=\epsilon_{\rm 0}\hat{{\mathbf E}}({\mathbf r})
+\hat{{\mathbf P}}({\mathbf r})$ we find for the wave equation
of the electric field in the rotating frame
\begin{align}
&\left[\Bigl(\partial_t+\Omega(t)\partial_\varphi
\Bigr)^2- \,c^2\,\bigtriangleup\right]
\hat{{\mathbf E}}({\mathbf r},t)\nonumber\\
&\qquad\qquad=\frac{1}{\epsilon_0}\Bigl(
\partial_t+\Omega(t)\partial_\varphi\Bigr)^2 \hat{{\mathbf P}}
({\mathbf r},t).
\end{align}
We now introduce slowly varying variables for the transverse field 
as well as polarization by
$\hat{{\mathbf E}}({\mathbf r},t)
=\hat{\pmb{\mathcal E}}^{(+)}(x,r_{\bot},t)\,e^{-{\rm i}(\omega_p t-k_p x)}+h.a.$ 
and $\hat{\mathbf P}({\mathbf r},t)
=\hat{\pmb{\mathcal P}}^{(+)}(x,r_{\bot},t)\,e^{-{\rm i}(\omega_p t-k_p x)}
+h.a.\,$, where $x=R\varphi$ is the arclength on the circle. 
Restricting ourselves to propagation along the periphery of the
interferometer we find within the slowly varying envelope approximation
and by neglecting terms ${\mathcal O}(\Omega\,R/c)$ 
\begin{equation}
\Bigl[\partial_t+c\partial_x+ {\rm i} k_{\rm p} \Omega R%-\frac{\dot\Omega\, R}{2 c}
\Bigr]\,
\hat{\pmb{\mathcal E}}^{(+)}(x,t)=-
\frac{{\rm i}\,\omega_{\rm p}}{2\,\epsilon_0}\hat{\pmb{\mathcal P}}^{(+)}(x,t).
\label{field-equation}
\end{equation}
The term proportional to the rotation rate $\Omega$ is responsible 
for the rotation induced Sagnac phase shift in the pure light case, 
i.~e.~without any influence from the medium polarization. 
As shown in \cite{Zimmer-PRL-2004} and in the next section 
the polarization leads to an additional phase shift. 

Introducing also slowly varying amplitudes for the matter fields
$\hat\Psi_1=\hat\Phi_1$, $\hat\Psi_2=\hat\Phi_2\,
e^{-{\rm i}(\omega_p t-k_p x)}$ and 
$\hat\Psi_3=\hat\Phi_3\,e^{-{\rm i}(\Delta\omega t-\Delta k x)}$ with 
$\Delta\omega=\omega_p-\omega_c$ and $\Delta k=k_p-k^\parallel_c$, where $k^\parallel_c$
is the wavevector projection of the control field onto the $x$-axis,
we find
\begin{align}
\Bigl({\mathcal D}_{\rm 1}-V_1(x)\Bigr)\hat{\Phi}_1 &= 
\hbar\Omega_p^*\hat\Phi_2,
\label{SloVarEqnSta1}\\
\Bigl({\mathcal D}_{\rm 2}+\hbar(\Delta_2-k_p\Omega R-V_2(x))
\Bigr)\hat{\Phi}_2 &= \hbar\Omega_p\hat\Phi_1\nonumber\\
&+\hbar\Omega_c\hat\Phi_3,
\label{SloVarEqnSta2}\\
\Bigl({\mathcal D}_{\rm 3}-V_3(x)+\hbar(\Delta_3-\eta k_p\Omega R
)\Bigr)\hat{\Phi}_3 &= \hbar\Omega_c^*\hat\Phi_2
\label{SloVarEqnSta3}
\end{align}
with 
\begin{equation}
{\mathcal D}_{\mu}={\rm i}\,\hbar\,\partial_t+\frac{\hbar^2\partial^2_x}
{2 m}+{\rm i}\,\hbar\,(\Omega R+\eta_\mu v_{\rm rec})\partial_x.
\end{equation}
Here we have used the definitions 
$\Delta_2=\omega_p-\omega_2-\omega_{\rm rec}$ 
and $\Delta_3=\Delta\omega-\omega_3-\eta^2{\omega_{\rm rec}}$ 
for the one- and two-photon-detuning including the recoil 
shift ($\omega_{\rm rec}=\hbar k_p^2/2m$). $v_{\rm rec}=\hbar k_p/m$ is the
single-photon recoil velocity.
We have also introduced the dimensionless parameter 
$\eta=\Delta k/k_p$ which describes the momentum transfer from 
the light fields to the atoms in state $|3\rangle$ as well as the abbreviation 
$\eta_\mu=\delta_{\mu,2}+\eta\,\delta_{\mu,3}$.
Finally the definitions $\Omega_{p,c}=-{\mathbf d}_{p,c}\cdot{\mathbf E}^{(p,c)}_{\rm ext}$  for the probe and control-field Rabi frequencies were applied. The shortened wave equation (\ref{field-equation}) and the matter-wave field equations (\ref{SloVarEqnSta1})-(\ref{SloVarEqnSta3})
are the basis of the following study of the sensitivity enhancement of the light--matter-wave hybrid Sagnac interferometer.
%%%%%%%%%%%%%%%%%%%%%%%%%%%%%%%%%%%%%%%%%%%%%%%%%%%%%%%%%%
%%%%%%%%%%%%%%%%%%%%%%%%%%%%%%%%%%%%%%%%%%%%%%%%%%%%%%%%%%%
%
\section{Sagnac phase shift and influence of external 
trapping potentials}
\label{Sec:InfExtTraPot}
%
%%%%%%%%%%%%%%%%%%%%%%%%%%%%%%%%%%%%%%%%%%%%%%%%%%%%%%%%%%
%%%%%%%%%%%%%%%%%%%%%%%%%%%%%%%%%%%%%%%%%%%%%%%%%%%%%%%%%%%
In this section we will calculate the stationary Sagnac phase shift 
for the hybrid interferometer 
in   the perturbative limit of low
probe-light intensities. 
In particular we will analyze the 
effects of  a trap potential which confines the atoms to certain regions in the direction of
the interferometer path. For simplicity we assume a constant rotation
rate, i.e. $\dot\Omega=0$, and consider the stationary state.
All atoms are assumed to be initially, i.~e.~before applying any probe-field, 
in the internal state $|1\rangle$. This amounts to set $\hat \Phi_2^{(0)}(t=0)=\hat 
\Phi_3^{(0)}(t=0)=0$.
In the perturbative limit of  small
probe field intensity, $\hat \Phi_1$ is not changed by the atom-light
interaction, i.~e.~it obeys the equation
\begin{equation}
\Bigl(\frac{\hbar^2\partial_x^2}{2 m} +{\rm i}\hbar \Omega R \partial_x+
\left(\epsilon_1-V_1(x)\right)\Bigr)
\hat\Phi_1^{(0)}(x)=0,\label{ZerothOrderPhi1}
\end{equation}
where $\epsilon_1$ is the energy in the stationary state.
Assuming $|\Omega_c| \gg |\Delta_2|, k_p |\Omega| R, |V_2(x)|/\hbar $ one finds
in first order from eq.~(\ref{SloVarEqnSta2})
\begin{equation}
\hat\Phi_3^{(1)}(x)=-\frac{\Omega_p(x)}{\Omega_c}\hat\Phi_1^{(0)}(x),
\label{AdiaElimEqu}
\end{equation}
which amounts to an adiabatic elimination of the excited state.
Using this and eq.~(\ref{SloVarEqnSta3}) we find
\begin{align}
\hat\Phi_2^{(1)}=&\,
\eta \,k_p \frac{\Omega R}{|\Omega_c|^2}\hat\Phi_1^{(0)}\Omega_p(x)
-\frac{1}{|\Omega_c|^2}\left[\frac{\hbar}{2m}\partial_x^2\right.\nonumber\\
&\hspace{-0.25cm}\left.+{\rm i}\,(\Omega R+\eta v_{\rm rec})\partial_x
+\frac{\epsilon_3-V_3(x)}{\hbar}
%+{\rm i}\gamma_3
\right]\hat\Phi_1^{(0)}\Omega_p(x)
\label{eq:FirsOrdeEquaStat2}\\
=&\enspace \eta\,\frac{\hat\Phi_1^{(0)}}{|\Omega_c|^2}
\Bigl({\rm i} v_{\rm rec}\partial_x \ln
{\hat\Phi}_1 +k_p\Omega R\Bigr)\Omega_p(x)\nonumber\\
-\frac{\hat\Phi_1^{(0)}}{|\Omega_c|^2}&\left[\frac{\hbar\partial_x^2}{2m}
+{\rm i}\Bigl(\Omega R+\eta v_{\rm rec} -{\rm  i} \frac{\hbar}{m}\partial_x 
\ln {\hat \Phi}_1^{(0)}\Bigr)\partial_x\right]\Omega_p(x),\nonumber
\end{align}
where we have in addition assumed two-photon resonance, i.~e.~$\Delta_3=0$.
In deriving the second equation, which is useful
for later discussions, we have made use of eq.~(\ref{ZerothOrderPhi1})
and assumed equal trapping potentials for the internal states
$V_1=V_3$. Furthermore an unimportant constant energy term
proportional to $\epsilon_1-\epsilon_3$ has been dropped.
One recognizes that the fields $\hat\Phi_2^{(1)}$ and $\hat\Phi_3^{(1)}$
and thus the medium polarization in first order of perturbation 
follow straight forwardly from the solution of eq.(\ref{ZerothOrderPhi1}).

We will now consider two cases. In the first case no confining 
potential for atoms in state $|1\rangle$ is assumed, which is equivalent to translational invariance on a ring. 
In the second case, discussed
later, a trapping potential in the longitudinal direction $x$ is taken into account. 
We will see that both cases lead to quite different results.
%%%%%%%%%%%%%%%%%%%%%%%%%%%%%%%%%%%%%%%%%%%%%%%%%%%%%%%%
%
\begin{figure}[tb]
\begin{center}
\includegraphics[width=0.45\textwidth]{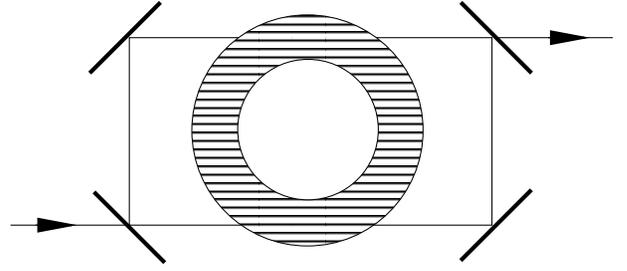}
\caption{Set-up of a Sagnac interferometer with ring-shaped trap configuration supporting a superfluid ultra-cold gas (BEC). The symmetric interferometer set-up allows for a distinction of rotational from linear acceleration. \label{fig:RingTrap}
}
\end{center}
\end{figure}
%
%%%%%%%%%%%%%%%%%%%%%%%%%%%%%%%%%%%%%%%%%%%%%%%%%%%%%%%%%%%5
%
%
%%%%%%%%%%%%%%%%%%%%%%%%%%%%%%%%%%%%%%%%%%%%%%%%%%%%%%%%%%%
%
\subsection{Periodic boundary conditions in state $|1\rangle$}
\begin{figure}[tb]
\begin{center}
\includegraphics[width=0.42\textwidth]{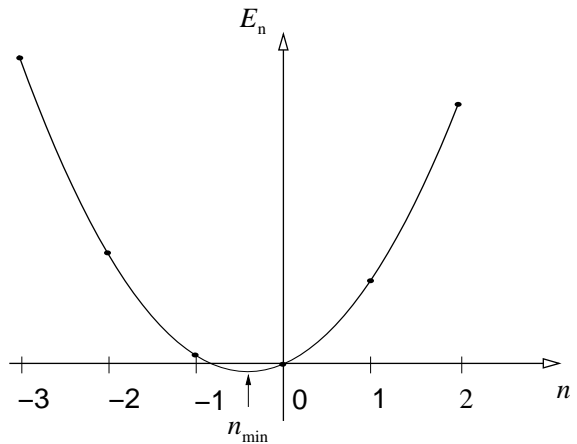}
\caption{Parabolic spectrum $\epsilon_n$ with minimum at $n_{\rm min}=-m \Omega R^2/\hbar$
when taken as a function of the continuous parameter $n$.}
 \label{fig:spectrum}
\end{center}
\end{figure}

%
%%%%%%%%%%%%%%%%%%%%%%%%%%%%%%%%%%%%%%%%%%%%%%%%%%%%%%%%%%%
Let us consider the case that atoms in state $|1\rangle$ do not
experience any confining potential in the $x$-direction. Since $x$ is
the coordinate along the periphery of the interferometer, this 
amounts to considering a ring-trap configuration with 
periodic boundary
conditions $\hat \Phi_1^{(0)}(x+2\pi R)=\hat\Phi_1^{(0)}(x)$. 
The principle set-up is shown in Fig.~(\ref{fig:RingTrap}).
With $V_1(x)\equiv 0$, eq.~(\ref{ZerothOrderPhi1}) has the eigensolutions
\begin{displaymath}
{\hat\Phi}_{1n}^{(0)}(x)= \hat\Phi_0\, \exp\left\{i \frac{n}{R}x\right\}
\qquad \epsilon_n=n\hbar\Omega +\frac{n^2\hbar^2}{2 m  R^2}
\end{displaymath} 
where ${\hat \Phi}_{0}$ is a constant. 
Taken as a continuous function of $n$, the spectrum $\epsilon(n)$ is a parabola with minimum at
\begin{equation}
n_{\rm min}= - \frac{m\Omega R^2}{\hbar}.
\end{equation}
This is illustrated in Fig.~\ref{fig:spectrum}.
Taking into account that $n$ must be a positive or negative integer, the state with the 
lowest energy corresponds to $n=0$
as long as $|n_{\rm min}|< 1/2$, i.e. as long as 
the Sagnac phase shift per round trip is smaller than $\pi$. 

%%%%%%%%%%%%%%%%%%%%%%%%%%%%%%%%%%%%%%%%%%%%%

\subsubsection{Bose-Einstein condensate}

%%%%%%%%%%%%%%%%%%%%%%%%%%%%%%%%%%%%%%%%%%%%%

We now assume that only the lowest motional energy state in the internal state $|1\rangle$ 
is initially excited, e.g. a Bose-Einstein condensate in the ring trap. In this case there
is a uniform phase over the whole ring and we can set $\hat\Phi_1^{(0)}(x)=\hat\Phi_0$.
It is important to note that the particles in the ground state do not attain a rotational phase shift
in this case.
This yields with eq.~(\ref{eq:FirsOrdeEquaStat2})
\begin{eqnarray}
&&\hat\Phi_2^{(1)} (x)= 
\eta k_p\frac{\Omega R}{|\Omega_c|^2} {\hat \Phi}_0\, \Omega_p(x) \\
&&\quad- \frac{{\rm i} (\Omega R +\eta v_{\rm rec}){\hat \Phi}_{0}}{|\Omega_c|^2} 
\partial_x\, \Omega_p(x) -\frac{1}{|\Omega_c|^2}{\hat \Phi}_0
\frac{\hbar}{2 m}\partial_x^2 \, \Omega_p(x).
\nonumber
\end{eqnarray}
Substituting the expressions for $\hat\Phi_2^{(1)}$ and $\hat\Phi_1^{(0)}$
into the stationary, shortened wave equation, eq.(\ref{field-equation}),
for the expectation value
of the probe-field expressed in terms of $\Omega_p$, 
\begin{equation}
\Bigl(c\partial_x + {\rm i}k_p \Omega R\Bigr)\Omega_p(x) = -{\rm i} g^2 \Bigl\langle
{\hat \Phi}_1^{(0)\dagger}(x){\hat \Phi}_2^{(1)}(x)\Bigr\rangle
\end{equation}
where $g=d_{12} \sqrt{\omega_p/2\hbar\epsilon_0 F}$, $d_{12}$ being the
dipole matrix element of the $|1\rangle \leftrightarrow |2\rangle$ transition
and $F$ the transversal cross-section of the probe-beam, we find
\begin{eqnarray}
&&\hspace{-0.25cm}\biggl[\bigl\{c\cos^2\theta + (\eta v_{\rm rec}+\Omega R)\sin^2\theta\bigr\}
\partial_x 
-{\rm i}\sin^2\theta \frac{\hbar\partial_x^2}{2 m}\biggr]\Omega_p(x)\nonumber\\
&&=
-{\rm i} k_p\Omega R\Bigl(\cos^2\theta +\eta \sin^2\theta\Bigr)\Omega_p(x).
\label{field-2}
\end{eqnarray}
Here we have introduced the mixing angle $\theta$ through
$\tan^2\theta=g^2 \varrho/|\Omega_c|^2$, where $\varrho = \langle \hat\Phi_0^\dagger
\hat\Phi_0\rangle$ is the density of atoms in state $|1\rangle$. 
Eq.~(\ref{field-2}) has a very intuitive interpretation. It describes
the propagation of the probe-field with the  group velocity
\begin{equation}
v_{\rm gr}=c\,\cos^2\theta + \eta\,v_{\rm rec}\, \sin^2\theta
\label{eqn:PolaritonGroupVelocity}
\end{equation}
in the rotating frame. 
The propagation of light in an EIT medium is associated with the
formation of a dark-state polariton, a superposition
of electromagnetic and matter-wave components \cite{Fleischhauer-PRL-2000}.
If we neglect the atomic motion, the group velocity of this
quasi-particle is proportional to the
square of the weight factor $\cos\theta$ of the electromagnetic part of the
polariton. However, if the coherence transfer from
light to atoms is accompanied by a finite
momentum transfer of $\eta\, m\, v_{\rm rec}$, there is also a 
matter-wave contribution to the total group velocity (\ref{eqn:PolaritonGroupVelocity}). 
This contribution is again proportional to the square of the
weight factor $\sin\theta$ of the matter-wave part.
Due to the admixture of the matter wave, the equation
of motion (\ref{field-2}) attains a term corresponding to the kinetic 
energy of this component
which leads to a dispersive spreading of the probe field along its 
propagation direction. This term becomes important
in the limit $\tan^2\theta >  \tan^2\theta_{\rm crit}\equiv c/v_{\rm rec}$, i.e. when the
light wave essentially turns into a propagating spin polarization.
The right hand side of eq.~(\ref{field-2}) describes the light and
matter-wave contributions to the rotational phase shift. 
The matter-wave contribution to the phase shift 
is non-zero only if there is a finite
momentum transfer, i.e. if $\eta \ne 0$. 
In the limit of small rotation,
$|\Omega|R \ll v_{\rm gr}$, which is the case of interest here, eq.~(\ref{field-2}) can easily be solved. 
Neglecting the second-order derivative the equation %eq.~(\ref{field-2})
reduces to eq.~(11) of ref.~\cite{Zimmer-PRL-2004}
\begin{equation}
\partial_x \ln \Omega_p(x)=- {\rm i} \frac{2\pi\Omega R}{\lambda c}
\left(\frac{\xi(x)}
{\xi(x)+\eta}
 + \frac{mc^2}{\hbar \omega_p}\frac{\eta}{\xi(x)+\eta}\right),
\label{field-3}
\end{equation}
where 
\begin{equation}
\xi(x)\equiv \frac{\cot^2\theta}{\cot^2\theta_{\rm crit}}
\approx \frac{v_{\rm gr}(x)}{v_{\rm rec}}-\eta.\label{eq:xi-def}
\end{equation}
The last approximate equation is only valid for $v_{\rm gr}\ll c$.
When $\xi$ is large the group velocity is much larger than the
recoil velocity, while $\xi$ approaching zero 
means that 
the group velocity is comparable to the recoil velocity.
Eq.~(\ref{field-3}) describes a phase shift of the probe field
in a medium without absorption, which is  canceled  due to  EIT. 
Hence two counter-propagating probe fields will experience the Sagnac phase shift
\begin{equation}
\Delta\phi= \frac{2\pi \Omega R}{\lambda c}
\int\!\! {\rm d}x\, \frac{\xi(x)}{\xi(x)+\eta}
+\frac{\Omega R}{\hbar/m}
\int\!\! {\rm d}x\, \frac{\eta}{\xi(x)+\eta}.
\end{equation}
This is the result obtained in \cite{Zimmer-PRL-2004}. It has two terms,
a light-contribution and, if $\eta\ne 0$, a matter-wave contribution.
Its most important consequence is that if the group velocity becomes comparable to the recoil velocity, i.e.
for $\xi \to 0$, the slow-light Sagnac phase approaches the matter-wave value!

%%%%%%%%%%%%%%%%%%%%%%%%%%%%%%%%%%%%%%%%%%%%%%%%%%%%%%%
%
%
%%%%%%%%%%%%%%%%%%%%%%%%%%%%%%%%%%%%%%%%%%%%%%%%%%%%%%%%%%%5
%
%

%%%%%%%%%%%%%%%%%%%%%%%%%%%%%%%%%%%%%%%%%%%%

%%%%%%%%%%%%%%%%%%%%%%%%%%%%%%%%%%%%%%%%%%%%%%%%%

\subsubsection{thermal gas}

%%%%%%%%%%%%%%%%%%%%%%%%%%%%%%%%%%%%%%%%%%%%%%%%%

The ground-state solution $\hat\Phi_1^{(0)}(x)=\hat\Phi_{1,n=0}^{(0)}=\hat\Phi_0=const.$
means that the atoms do not follow the
motion of the trap. This is strictly speaking only possible if the
gas is superfluid. In a normal gas collisions with wall roughnesses and
between atoms, which are not taken into
account here, would accelerate the vapor particles in the initial phase of
rotation. Eventually an equilibrium state would be reached where
the atoms co-rotate with the trap. This can also be seen from a different
argument. In a thermal state with $k_B T \gg \hbar\Omega +\hbar^2 /2 m R^2$
many states in the spectrum of Fig.~\ref{fig:spectrum} will be occupied. As a
consequence the thermal gas in the ground state attains an average rotational phase
($x=2\pi R$)
\begin{equation}
\overline{\Delta\phi}= - 2\pi \langle n\rangle \to  -2\pi\, n_{\rm min}= \frac{2\pi \Omega R^2}{\hbar/m}. 
\end{equation}
This is just the matter-wave Sagnac phase and is in sharp contrast to the case of
a Bose-Einstein condensate, where the ground state does not acquire any rotational phase. 
Since now both, the ground state $|1\rangle$ {\it and} the excited state $|2\rangle$ attain
the same Sagnac phase shift, the matter-wave contribution to the polarization is exactly cancelled.
Thus the extention to thermal gases made 
in \cite{Zimmer-PRL-2004} is not correct.

\

The need for a superfluid gas (e.g. BEC)
in a ring trap puts restrictions to the achievable interferometer area.
Although recently there has been substantial progress in realizing
ring traps for BEC \cite{Gupta-PRL-2005,Arnold-PRA-2006}, the area achieved is only on the
order of 10$^{-1}$ cm$^2$, which cannot compete with the values
reached in fiber-optical gyroscopes. 
%
%
%%%%%%%%%%%%%%%%%%%%%%%%%%%%%%%%%%%%%%%%%%%%%%%%%%%%%%%%%%%

\subsection{Effect of longitudinal confinement}

%%%%%%%%%%%%%%%%%%%%%%%%%%%%%%%%%%%%%%%%%%%%%%%%%%%%%%%%%%%
%
%
Let us now discuss the case of a longitudinal trapping potential
for atoms in state $|1\rangle$, i.e. $V_1(x)\ne 0$ in 
eq.~(\ref{ZerothOrderPhi1}). In this case the substitution 
\begin{equation}
{\hat\Phi}_1^{(0)}(x)={\hat \Phi}_0\, f(x)\, {\rm e}^{-{\rm i} m \Omega R x/\hbar}
\label{Phi1-with-potential}
\end{equation}
leads to the steady-state equation
\begin{equation}
\left(\frac{\hbar^2\partial_x^2}{2 m} +\frac{m}{2}\Omega^2 R^2
+\epsilon_1-V_1(x)\right) {f}(x) =0.
\end{equation}
If one disregards the small centrifugal energy shift proportional to
$\Omega^2$, this equation is just the stationary Schr\"odinger 
equation for a particle in the trap potential $V_1$. 
The solution of this equation is independent of the rotation rate $\Omega$.
(The inclusion of the centrifugal term would lead to a higher order contribution to
the Sagnac phase, which we are not interested in.)
If we substitute (\ref{Phi1-with-potential}) into the second equation
of (\ref{eq:FirsOrdeEquaStat2}), one recognizes that all terms containing
the rotation rate $\Omega$ vanish exactly:
\begin{eqnarray}
\hat\Phi_2^{(1)} &=&\hspace{-0.15cm} 
-{\rm i} \frac{\hat\Phi_1^{(0)}(x)}{|\Omega_c|^2}\left(\eta v_{\rm rec}
(\partial_x \ln f(x)) \Omega_p(x) -{\rm i} 
\frac{\hbar\partial_x^2}{2 m} \Omega_p(x)\right)\nonumber\\
&& \hspace{-0.15cm}-{\rm  i}\frac{\hat\Phi_1^{(0)}(x)}{|\Omega_c|^2}\left(\eta v_{\rm rec}
-{\rm i}\frac{\hbar}{m}\partial_x \ln f(x)\right)\partial_x\Omega_p(x).
\end{eqnarray}
Substituting this into the shortened wave equation for $\Omega_p$ yields
\begin{eqnarray}
\Bigl[c\cos^2\theta + \Bigl(\eta v_{\rm rec}
&-&{\rm i}\frac{\hbar}{m}\partial_x \ln f(x)\Bigr)\sin^2\theta\Bigr]\partial_x \Omega_p(x)
\nonumber\\
-{\rm i}\sin^2\theta \frac{\hbar\partial_x^2}{2 m}\Omega_p(x) &=&
-{\rm i}k_p\Omega R\, \cos^2\theta\, \Omega_p(x) 
\label{field-4}\\
&+&\eta v_{\rm rec} \sin^2\theta 
\Bigl(\partial_x \ln f(x)\Bigr) \Omega_p(x).
\nonumber
\end{eqnarray}
Neglecting the term with the second order derivative as well as those containing $\partial_x f(x)$,
which amounts to assume that 
$f(x)$ is a slowly-varying ground-state wave function of a sufficiently 
smooth potential, eq.(\ref{field-4}) reduces to
\begin{align}
\partial_x \ln \Omega_p(x)&=- {\rm i} \frac{2\pi\Omega R}{\lambda c}
\frac{\cos^2\theta}{\cos^2\theta+\eta\, v_{\rm rec}\sin^2\theta}\nonumber\\
&=
-{\rm i}\frac{2\pi\Omega R}{\lambda c}
\frac{1}{1+\eta\,c/\xi},
\label{field-5}
\end{align} 
It is
immediately obvious that only the light part of the Sagnac phase survives.
Thus in the EIT hybrid gyroscope a matter-wave contribution to the 
Sagnac phase only emerges in the absence of a confining potential or if 
periodic boundary conditions apply as e.g. in a ring trap. 

The physical interpretation of this result is straight forward. In the presence of a confining
potential the atoms in state $|1\rangle$ are bound to the motion of the
confining potential. Hence they acquire a rotational phase shift by following the
motion of the trap attached to the rotating frame \cite{Hendriks-QOPT-1990}. Atoms in state $|2\rangle$ acquire the same phase shift since they are in the same frame. Therefore, the polarization attains no Sagnac
phase as it is a sesqilinear function of the wave-functions of states $|1\rangle$ and
$|2\rangle$. This is in contrast to a superfluid  BEC in a ring trap, where the order parameter does 
not acquire any 
phase due to the periodic boundary conditions as long as the rotation is sufficiently slow.
%
%
%%%%%%%%%%%%%%%%%%%%%%%%%%%%%%%%%%%%%%%%%%%%%%%%%%%%%%%%%%%%%%%%%%%%%
%%%%%%%%%%%%%%%%%%%%%%%%%%%%%%%%%%%%%%%%%%%%%%%%%%%%%%%%%%
%
%
\section{Quantum limited sensitivity of the slow-light gyroscope}
%
%
%%%%%%%%%%%%%%%%%%%%%%%%%%%%%%%%%%%%%%%%%%%%%%%%%%%%%%%%%%%%%%%%%%%%%
%%%%%%%%%%%%%%%%%%%%%%%%%%%%%%%%%%%%%%%%%%%%%%%%%%%%%%%%%%
%
%
We now want to calculate the sensitivity of the 
slow-light Sagnac interferometer in the case of periodic
boundary conditions, i.e. in the absence of any confining
potential in the propagation direction. For simplicity we consider
the case $\eta=1$, i.~e.~perpendicular propagation directions
of probe and control field. 

To determine the sensitivity we assume that 
the error in determining the Sagnac phase is entirely 
given by shot-noise quantum fluctuations.
If coherent laser light or Poissonian
particle sources are used the shot-noise limit of the
phase measurement is given by
\begin{equation}
\Delta\phi_{\rm noise} = \frac{1}{\sqrt{n_{\rm D}}},
\label{shot-noise}
\end{equation}
where $n_{\rm D}=I_{\rm out} t_{\rm D}$ is the total number of 
photons or atoms
counted at the detector during the measurement time $t_{\rm D}$ \cite{Scully-1997}.
Here $I_{\rm out}$ is the photon or atom flux.
The assumption that the quantum noise limit is set by shot-noise
is justified by two observations:
First of all, it is well known that
using nonclassical light or sub-Poissonian particle sources
does in general not lead to an improvement of the signal-to-noise
ratio  in
interferometry since at the optimum operation point the amplitude reduction due to 
losses is typically of order ${\rm e}^{-1}$ and thus quite substantial.
These losses tend to quickly 
destroy the fragile nonclassical and sub-Poissonian properties.
Secondly, as has been shown in 
\cite{Scully-PRL-1992,Fleischhauer-PRA-1992}, atomic noise
contributions in EIT-type interferometer set-ups are small and can be 
neglected. 

\begin{figure}[tb]
\begin{picture}(80,140)(0,0)
\put(-62,8){\large $|1\rangle$}
\put(-6,117){\large $|2\rangle$}
\put(120,36){\large $|3\rangle$}
\put(18,70){\large $\Omega_p$}
\put(50,70){\large $\Omega_c$}
\put(32,10){\large $\gamma_{\rm 13}$}
\put(-17,70){\large $\gamma_{\rm 1}$}
\put(87,70){\large $\gamma_{\rm 3}$}
\put(-50,10){\includegraphics[width=6cm]{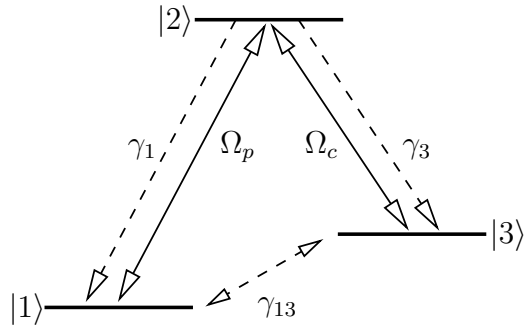}}
\end{picture}
\caption{$\Lambda$ configuration in which the Rabi frequency 
$\Omega_p$ drives the $1\leftrightarrow 2$-transition and 
$\Omega_c$ the $3\leftrightarrow2$-transition (solid lines). 
Radiative decay from the excited level to $|1\rangle$ or to 
$|3\rangle$ goes as $\gamma_{\rm 1}$ or $\gamma_{\rm 3}$ 
respectively (dashed lines). The dephasing rate of the $1-3$ 
coherence is denoted by $\gamma_{13}$. }
\label{fig:LamSchWitDec}
\end{figure}

In the weak-signal limit discussed in the previous section, the Sagnac
phase accumulated is independent of the signal field strength 
\cite{Zimmer-PRL-2004}, hence the signal-to-noise ratio could become 
arbitrarily large when the input-laser power is increased. In reality
the Sagnac phase approaches a maximum value at a certain optimum
probe-laser power and decreases for larger intensities. The optimum
intensity is reached when the number density of photons in the EIT
medium approaches that of the atoms. Thus in order to calculate the 
maximum sensitivity and to find optimum operation conditions
we have to calculate the Sagnac phase to all orders of the signal Rabi 
frequency $\Omega_p$. Since in higher order perturbation the excited state $|3\rangle$ 
attains a finite population, decay out of the excited state needs to be taken
into account. The decay leads to a population redistribution among the
states of the $\Lambda$ system, see Fig.~\ref{fig:LamSchWitDec}. 
It can also lead to loss out of the system. 
We will disregard the latter process however. Furthermore, we assume that
the density of the considered medium is low enough that it is sufficent to describe
the system by a set of equations for the single-particle
density matrix elements
$ \rho_{\mu\nu}(x,x',t)=\langle{\hat\Phi}^\dagger_\nu(x',t)
{\hat\Phi}_\mu(x,t)\rangle $. Here $\mu,\nu\in\{1,2,3\}$ denote the internal states.
Since the medium polarization is 
determined by the local density-matrix element $\rho_{12}(x,x,t)$, i.e. $x^\prime=x$,
we consider only local quantities.
For the density matrix elements diagonal 
in the internal states we find the equations of motion
\begin{widetext}
\begin{align}
\partial_t\rho_{\rm 11}(x,t)&=\gamma_1\rho_{\rm 22}(x,t)-{\rm i}\,
\Omega_{\rm p}^* (x,t)\rho_{\rm 21}(x,t)+{\rm i}\,\Omega_{\rm p}(x,t) 
\rho_{\rm 12}(x,t)
%\nonumber\\
%&
+\Omega R\,\partial_x\rho_{\rm 11}(x,t),
\label{eq:vonNeumann1}\\[0.25cm]
\partial_t\rho_{\rm 22}(x,t)&=-\gamma_{2}\rho_{\rm 22}(x,t)+{\rm i}\,
\Omega_{\rm p}^* (x,t)\rho_{\rm 21}(x,t)-{\rm i}\,\Omega_{\rm p}(x,t) 
\rho_{\rm 12}(x,t)
\nonumber\\
&
+{\rm i}\,\Omega_{\rm c}^* (x,t)\rho_{\rm 23}(x,t)-{\rm i}\,\Omega_{\rm c}(x,t) \rho_{\rm 32}(x,t)+(\Omega R+v_{\rm rec})
\partial_x \rho_{\rm 22}(x,t),
\label{eq:vonNeumann2}\\[0.25cm]
% \end{align}
% %%
% \begin{align}
\partial_t\rho_{\rm 33}(x,t)&=\gamma_3\rho_{\rm 22}(x,t)-{\rm i}\,
\Omega_{\rm c}^*(x,t) \rho_{\rm 23}(x,t)+{\rm i}\,\Omega_{\rm c}(x,t) 
\rho_{\rm 32}(x,t)
\nonumber\\
&
+(\Omega R+ v_{\rm rec})
\partial_x\rho_{\rm 33}(x,t).
\label{eq:vonNeumann3}
\nonumber\\
\end{align}
Likewise we find for the local coherences%}\nonumber\\
\begin{align}
\partial_t\rho_{\rm 12}(x,t)&=-({\rm i}(\Delta_2+\Omega R k_{\rm p})+\gamma_2/2)
\rho_{\rm 12}(x,t)+{\rm i}\,\Omega_{\rm c}^*(x,t)\rho_{\rm 13}(x,t)\nonumber\\
&-{\rm i}\,\Omega_{\rm p}^*(x,t)(\rho_{\rm 22}(x,t)
-\rho_{\rm 11}(x,t))
+(\Omega R+v_{\rm rec})\partial_x\rho_{\rm 12}(x,t)\nonumber\\
&+v_{\rm rec}\langle\hat\Phi_{\rm 2}^\dagger(\partial_x\hat\Phi_{\rm 1})
\rangle\, , 
\label{eq:vonNeumann4}
\\[0.25 cm]
\partial_t\rho_{\rm 13}(x,t)&=-({\rm i}(\Delta_3+\Omega R k_p )
+\gamma_{13})\rho_{\rm 13}(x,t)-{\rm i}\,\Omega_{\rm p}^*(x,t)
\rho_{\rm 23}(x,t)\nonumber\\
&+{\rm i}\,\Omega_{\rm c}(x,t)\rho_{\rm 12}(x,t)
+(\Omega R+ v_{\rm rec})\partial_{x}\rho_{\rm 13}(x,t)
+ v_{\rm rec}\langle \hat\Phi_{\rm 3}^\dagger(\partial_x\hat\Phi_{\rm 1})
\rangle\, ,
\label{eq:vonNeumann5}
\\[0.25cm]
\partial_t\rho_{\rm 23}(x,t)&={\rm i}(\Delta_2-\Delta_3)
-\gamma_2/2)\rho_{\rm 23}(x,t)-{\rm i}\,\Omega_{\rm p}(x,t)
\rho_{\rm 13}(x,t)\nonumber\\
&-{\rm i}\,\Omega_{\rm c}(x,t)(\rho_{\rm 33}(x,t)
-\rho_{\rm 22}(x,t))+(\Omega R+
v_{\rm rec})\partial_x
\rho_{\rm 23}(x,t).
\label{eq:vonNeumann6}
\end{align}
\end{widetext}
% 
% %
%
where $\gamma_2\equiv\gamma_1+\gamma_3$. In the following we determine the Sagnac phase shift
for arbitrary probe-field Rabi frequency based on the above set of equations and the shortened wave equation.
To derive a transparent expression for the rotationally induced phase shift further simplifications are however necessary.
%
%
%%%%%%%%%%%%%%%%%%%%%%%%%%%%%%%%%%%%%%%%%%%%%%%%%%%%%%%%%%%
%
\subsection{Non-local terms}
%
%%%%%%%%%%%%%%%%%%%%%%%%%%%%%%%%%%%%%%%%%%%%%%%%%%%%%%%%%%%
%
%
One recognizes from eq.~(\ref{eq:vonNeumann4}) and (\ref{eq:vonNeumann5}) that the local 
off-diagonal matrix elements are coupled to non-local quantities
of the form  $\langle \hat\Phi^\dagger_{\rm \mu}(x)
(\partial_x\hat\Phi_{\rm \nu}(x))\rangle$. These terms cause the build-up
of coherences between different internal states {\it and} different positions, which are zero in lowest-order 
perturbation. 
We now want to argue that these terms can be neglected. To this end we consider eq.~(\ref{eq:FirsOrdeEquaStat2}) again
disregarding second-order derivates and set $V_3\equiv\epsilon_3\equiv0$. Hence 
we have
\begin{align}
\hat\Phi_2=\frac{k_p\Omega
R}{|\Omega_c|^2}\hat\Phi_1\Omega_p
-{\rm i}
\frac{\Omega
R+v_{\rm
rec}}{|\Omega_c|^2}\partial_x(\hat\Phi_1\Omega_p).
\end{align}
Substituting this into the steady-state version of the equation of motion for $\hat\Phi_1$, eq.~(\ref{SloVarEqnSta1}), remembering
that there is no confining potential for atoms in state $|1\rangle$ in the propagation direction, we find
\begin{equation}
\partial_x\hat\Phi_1(x)=-{\rm i}s
\frac{\bigl[
k_p \Omega R-{\rm i}(\Omega R+v_{\rm rec})
(\partial_x\ln\Omega_p)
\bigr]}{\Omega R(1+s)+v_{\rm rec}s}
\hat\Phi_1(x),
\end{equation}
where $s=|\Omega_p|^2/|\Omega_c|^2$ is a saturation parameter. Since the probe field picks 
up a Sagnac phase shift, we have 
\begin{equation}
\partial_x\ln\Omega_p\sim-{\rm i}\frac{\alpha}{c}k_p
\Omega R.
\end{equation}
With the help of this we finally arrive at
\begin{equation}
\partial_x\hat\Phi_1=-{\rm i}\frac{k_p\Omega R}{v_{\rm
rec}}\left(1-\alpha\frac{v_{\rm rec}}{c}\right)
\hat\Phi_1+\mathcal{O}\Bigl((\Omega R)^2\Bigr).
\end{equation}
As a consequence the term 
$v_{\rm rec}\langle\hat \Phi^\dagger_2\partial_x\hat \Phi_1\rangle $
in eq.~(\ref{eq:vonNeumann4}) is of the order of 
\begin{equation}
v_{\rm rec}\langle\Phi^\dagger_2\partial_x\Phi_1\rangle  
\simeq
-{\rm i}k_p\Omega R
\left(1-\alpha\frac{v_{\rm rec}}{c}\right)
\rho_{12}\label{eq:nondiagonal-12}
\end{equation}
and is thus negligible as compared to
$\gamma_2\rho_{12}/2$. Using similar arguments one finds that the term $v_{\rm rec}\langle\Phi^\dagger_3\partial_x\Phi_1\rangle$ in eq.~(\ref{eq:vonNeumann5}) 
is of the order of 
\begin{equation}
v_{\rm rec}\langle\Phi^\dagger_3\partial_x\Phi_1\rangle  
\simeq
-{\rm i}k_p\Omega R
\left(1-\alpha\frac{v_{\rm rec}}{c}\right)
\rho_{13}.
\end{equation}
Since ideally the ground-state coherence is long-lived,
one has $\gamma_{13}\to0$. Hence neglecting this
term is not as straight forward as for eq.(\ref{eq:nondiagonal-12}). However,
adiabatically eliminating the fast decaying
optical coherence $\rho_{12}$ in eq.~(\ref{eq:vonNeumann4}) 
and substituting the resulting expression into the equation of motion of $\rho_{13}$, eq.~(\ref{eq:vonNeumann5}),
yields a term proportional to
$|\Omega_c|^2\gamma_{2}\rho_{13}/2$ which is much
larger than $k_p\Omega R\rho_{13}$. Thus also the term
$v_{\rm gr}\langle \hat\Phi_3^\dagger \partial_x\hat \Phi_1\rangle$
 can be safely neglected.
As a result of this approximation the density matrix equations 
(\ref{eq:vonNeumann1})-(\ref{eq:vonNeumann6})
are self-contained and local.
%
%%%%%%%%%%%%%%%%%%%%%%%%%%%%%%%%%%%%%%%%%%%%%%%%%%%%%%%%%%%%%%%%%%
%%%%%%%%%%%%%%%%%%%%%%%%%%%%%%%%%%%%%%%%%%%%%%%%%%%%%%%%%%%%%%%%%%
%
%
%
%%%%%%%%%%%%%%%%%%%%%%%%%%%%%%%%%%%%%%%%%%%%%%%%%%%%%%%%%%%
%
\subsection{Perturbation theory with respect to characteristic length}
%
%%%%%%%%%%%%%%%%%%%%%%%%%%%%%%%%%%%%%%%%%%%%%%%%%%%%%%%%%%%
%
%
In the following we assume one- and two-photon resonance, 
i.e.~$\Delta_2=\Delta_3=0$, and solve the above system of 
equations for the coherence of the $|1\rangle \leftrightarrow |2\rangle $-transition 
in steady state to all orders in $\Omega_p$.
The density matrix equations (\ref{eq:vonNeumann1})-(\ref{eq:vonNeumann6}) 
can, neglecting terms proportional to $\Omega R\partial_x$, be written in compact form as
\begin{equation}
\dot\rho(x,t)=\Bigl({\mathbf M}(x)+v_{\rm rec}{\mathbf D}\partial_x\Bigr)\rho(x,t),
\label{matrix-equation}
\end{equation}
where ${\mathbf M}$ and ${\mathbf D}$ are $9\times 9$ matrices. 
Even under stationary conditions we are still left with
a set of first order linear differential equations with space dependent
coefficient. Thus in order to find an analytic solution 
further approximations are needed.
To this end, we make use of the fact that the  
off-diagonal density matrix elements are only slowly varying in space.
Let $L$ and $T$ be their characteristic length and time scales.
Normalizing time and space to these units by $\xi=x/L$ and
$\tau = t/T$, eq.(\ref{matrix-equation}) reads
\begin{equation}
\partial_\tau \rho = 
\left(\tilde{\mathbf M}+\frac{v_{\rm rec}T}{L} 
\tilde{\mathbf D}\,\partial_\xi\right)\rho
\end{equation}
where typical matrix elements of $\tilde{\mathbf M}={\mathbf M}T$ read as $k_p R \Omega T$, 
with $k_p R,|\Omega T|\gg 1$
and those of $\tilde{\mathbf D}={\mathbf D}T$ are of order unity. Since 
$v_{\rm rec}T/L$ is typically small compared to unity
we can apply a perturbation expansion in this parameter. 

In zeroth order
we disregard the term containing ${\mathbf D}$. 
Hence in steady state we have to solve ${\mathbf M}\rho^{(0)}_{\rm ss}=0$ 
with the 
constraint $\sum_\mu \rho_{\mu\mu}(x)=n(x)$, which reflects the conservation
of probability.

In first order we find
\begin{equation}
\underline{\rho}_{\rm ss}^{(1)}(x)=\Bigl(\underline{{\mathbf 1}}-
v_{\rm rec}\underline{{\mathbf M}}^{-1}\underline{{\mathbf D}}
\partial_x\Bigr)\,
\underline{\rho}_{\rm ss}^{(0)}(x).
\label{eq:CharLenPerTheFirstOrder}
\end{equation}
Here $\underline{\mathbf M}$ is a reduced $8 \times 8$ matrix obtained
from ${\mathbf M}$ by incorporating the constraint 
$\sum_\mu \rho_{\mu\mu}(x)=n(x)$ and $\underline{\rho}_{\rm ss}^{(0)}$
is the corresponding zeroth-order density matrix.
The explicit expressions of all matrices and vectors 
can be
obtained from (\ref{eq:vonNeumann1})-(\ref{eq:vonNeumann6}) 
in a straight forward manner. They are however lengthy
and will not be given here. 

%%%%%%%%%%%%%%%%%%%%%%%%%%%%%%%%%%%%%%%%%%%%%%%%%%%%%%%%%%%%
%
\subsection{Steady state Maxwell-Bloch equation}
%
%%%%%%%%%%%%%%%%%%%%%%%%%%%%%%%%%%%%%%%%%%%%%%%%%%%%%%%%%%%%
%
To obtain the rotationally induced phase shift we expand 
eq.~(\ref{eq:CharLenPerTheFirstOrder}) up to first order 
in the angular velocity $\Omega$ and use the time-independent 
Maxwell equation (\ref{field-equation}) in the rotating frame 
\begin{equation}
\Bigl(c\partial_x-{\rm i}k_{\rm p}\Omega R\Bigr)\Omega_{\rm p}(z)
=-{\rm i}\,g^2 N \rho_{\rm 21}^{\rm ss}.
\label{eqn:MaxBloEqnProbFiel}
\end{equation}
To determine $\rho_{\rm 21}^{\rm ss}$  we furthermore neglected terms ${\mathcal O}(\gamma_{\rm 13}^2)$ and 
$\gamma_{\rm 13}\Omega_{\rm p}^n$ with $n\in{\mathbb N}$ since we 
assume a long-lived coherence between the two lower states $|1\rangle$ 
and $|3\rangle$. In addition 
to this we made use of the EIT condition $\Omega_{\rm c}^2
\gg\gamma_{\rm 13}\gamma_{\rm 1}$ \cite{Fleischhauer-RMP-2005} and assumed
for simplicity $\gamma_1=\gamma_3=\gamma$. 

With these assumptions we arrive at the following expressions 
for the real and imaginary part of the susceptibility, which determine 
the dispersion and absorption of the medium
\begin{align}
\chi^{\,\prime}(\Omega_{\rm p})=\beta^{-1}\,
\frac{\displaystyle \Omega\,R}{\displaystyle c}\left(1+ g^2 N
\frac{\displaystyle\Omega_{\rm c}^2}{\displaystyle(\Omega_{\rm c}^2
+|\Omega_{\rm p}|^2)^2}\right)
\label{eqn:SusceptibilityReal}
\end{align}
\begin{align}
\chi^{\,\prime\prime}(\Omega_{\rm p})=-\beta^{-1}
\frac{\displaystyle\gamma_{\rm 13}}{\displaystyle k_p c}g^2 N
\frac{\displaystyle
%g^2 N
\Omega_{\rm c}^2}{\displaystyle(\Omega_{\rm c}^2+|\Omega_{\rm p}|^2)^2}
\label{eqn:SusceptibilityImaginary}
\end{align}
with
\begin{align}
\beta(\Omega_{\rm p})=\displaystyle 1+ 
\frac{\displaystyle v_{\rm rec}}{\displaystyle c}g^2\,
N\frac{\Omega_{\rm c}^4}
{(\Omega_{\rm c}^2+|\Omega_{\rm p}|^2)^3}.
\end{align}
The imaginary part of the complex susceptibility $\chi(\Omega_{\rm p})
=\chi^{\,\prime}+{\rm i}\,\chi^{\,\prime\prime}$ can be further simplified. 
One can easily see that the absorption constant
is bounded from above by 
\begin{equation}
\kappa= k_p\chi^{\prime\prime}\le \frac{\gamma_{13}}{c}\frac{g^2 N}
{\Omega^2_{\rm c}}=\frac{\gamma_{13}}{v_{\rm rec}}\xi^{-1}.
\end{equation}
In this limiting case the following equation arises
\begin{align}
\partial_z \ln\Omega_{\rm p}(z)=
-\frac{\gamma_{13}}{c}\tan^2\theta+{\rm i}\,k_p\,\chi^{\,\prime}(\Omega_{\rm p}).
\label{eqn:SagnacPhaseEqua}
\end{align}
The first term in eq.~(\ref{eqn:SagnacPhaseEqua}) describes absorption losses
due to the nonvanishing decay of the ground-state coherence,
the second term the rotationally induced or Sagnac phase. 
Since the saturation of the absorption for
increasing probe-field intensities is not taken into account, the losses are slightly
overestimated.
%
%
%%%%%%%%%%%%%%%%%%%%%%%%%%%%%%%%%%%%%%%%%%%%%%%%%%%%%%%%%%%
%
\subsection{Quantum limit of gyroscope sensitivity}
\begin{figure}[th]
\includegraphics[width=7.5cm]{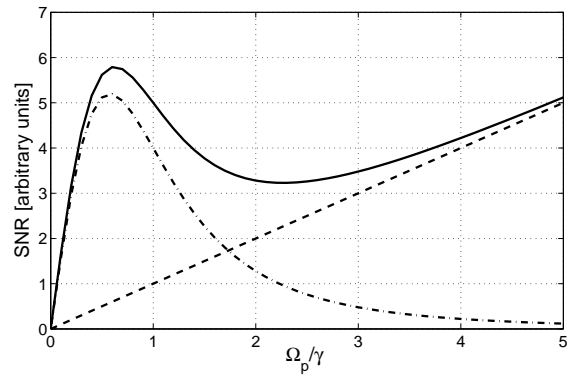}
\caption{Schematic dependence of SNR on input probe-field
Rabi frequency. The dash-dotted line indicates the contribution 
of the matter-wave term, the dashed line that of the light term.
The solid line is the sum of both contributions.}
\label{fig-3}
\end{figure}

%
%%%%%%%%%%%%%%%%%%%%%%%%%%%%%%%%%%%%%%%%%%%%%%%%%%%%%%%%%%%
%
%
Solving the shortened Maxwell equation (\ref{eqn:SagnacPhaseEqua}) for the probe field with the 
all-order susceptibility, eq.(\ref{eqn:SusceptibilityReal}), we can now determine 
the minimum detectable rotation rate $\Omega_{\rm min}$ 
of the slow-light gyroscope. 
This is done by maximizing the signal-to-noise 
ratio (SNR) of the interferometer with respect to the system parameters
and set it equal to unity.
The relative rotational phase shift of two 
polaritons propagating in opposite directions
is given by
\begin{equation}
\Delta\phi_{\rm sig} = \int\!{\rm d}x\, k_p\Bigl[
\chi^\prime\bigl(\Omega,\Omega_p(x)\bigr) 
- \chi^\prime\bigl(-\Omega,\Omega_p(x)\bigr)\Bigr].
\end{equation}
Using this and eq.~(\ref{eqn:SusceptibilityReal}) we find
\begin{eqnarray}
\Delta\phi_{\rm sig} &=& \frac{2\pi \Omega R}{\lambda c}\int\!{\rm d}x\, 
\frac{\xi(x)}{\xi(x)+\frac{1}{(1+s(x))^3}} +\nonumber\\
&& + \frac{\Omega R}{\hbar/m}\int\!{\rm d}x\, 
\frac{\frac{1}{(1+s(x))^2}}{\xi(x)+ \frac{1}{(1+s(x))^3}},\label{eq:signal-saturation}
\end{eqnarray}
where $s(x)=|\Omega_p(x)|^2/\Omega_c^2$ is the saturation parameter introduced before,
and $\xi(x)$, defined in eq.~(\ref{eq:xi-def}), determines the character of the polariton. 
One recognizes that the matter-wave part of the 
signal phase - the second line of eq.~(\ref{eq:signal-saturation}) -
decreases for increasing input probe intensity. The light part 
- first term in  eq.~(\ref{eq:signal-saturation}) - approaches  a
constant value in this limit. At the same time the shot-noise
phase error
\begin{equation}
\Delta\phi_{\rm noise} =\frac{1}{\sqrt{n_D}}\label{Delta-phi-noise}
\end{equation}
is inversely proportional to $|\Omega_p(0)|\,\exp\left(-\kappa L_M\right)$, where $L_M$ is the length of the medium and the
probe-field's source is located at $x=0$.
As a consequence of the different dependence of the signal and noise terms on the probe field strength,
the signal-to-noise ratio SNR $= \Delta\phi_{\rm sig}/
\Delta\phi_{\rm noise}$ has the qualitative behavior
shown in Fig.~\ref{fig-3}.
%
%
%%%%%%%%%%%%%%%%%%%%%%%%%%%%%%%%%%%%%%%%%%%%%%%%%%%%%%%%%%%%%%%%%%%%%%%
%%%%%%%%%%%%%%%%%%%%%%%%%%%%%%%%%%%%%%%%%%%%%%%%%%%%%%%%%%%%%%%%%%%%%%%%%
%
%
For very large laser fields the SNR becomes arbitrarily large, as the
light contribution to the Sagnac phase becomes intensity independent and
the shot-noise level decreases steadily. For small probe intensities the SNR
has a local maximum due to the saturation of the matter-wave phase shift.
As the matter-wave contribution to the Sagnac phase is orders of magnitude
larger than the light contribution, extremely large input intensities
would be required to exceed the sensitivity at the first local maximum.
(Note that Fig.~\ref{fig-3} is not drawn to scale.)
We thus consider only this first maximum when determining the 
quantum-limited sensitivity of the slow-light gyroscope. 

Although it is rather straight forward to calculate numerically, on the basis of  the
above given equations the minimum
detectable rotational phase shift, we
are interested here in an analytic estimate.
For this we make some simplifying assumptions: First of all we consider the propagation of polaritons
through a homogeneous medium. We furthermore ignore the space
dependence of the functions $\xi(x)$ and $s(x)$ in the expression
(\ref{eq:signal-saturation}) for the signal phase, which amounts to replacing $|\Omega_p(x)|$
by its input value $|\Omega_p(0)|\equiv |\Omega_p|$. As will be seen
this only slightly overestimates the saturation of the signal at the
optimum operation point. We also ignore the saturation of the probe field
absorption, which again merely slightly overestimates the probe field
losses at the operation point. Finally we only consider the dominant
matter-wave contribution to the signal phase.
Thus we have
\begin{equation}
\Delta\phi_{\rm sig} =\frac{\Omega R L}{\hbar/m} \frac{(1+s)}
{\xi (1+s)^3 +1}.
\end{equation}
In order to estimate the signal-to-noise ratio 
SNR $= \Delta\phi_{\rm sig}/\Delta\phi_{\rm noise}$ we now express the
the shot-noise expression (\ref{Delta-phi-noise}) in terms of the parameters
$\xi$ and $s$. The number of probe photons at the detector can be written 
in terms of the probe-field Rabi frequency at the source via
\begin{equation}
n_{\rm D}=\frac{P_{\rm D} t}{\hbar\omega_{\rm p}}
=\frac{2\,\epsilon_0\,F\,c}{\hbar\omega_p}\left(\frac{\hbar\,\Omega_p(0)}{|{\mathbf d}_p|}\right)^2
\,t\,e^{-2 \kappa L_M},
% =\frac{2}{3\pi}
% \frac{A_{\rm b}\omega_p^2}{c^2 \gamma}
% \Omega^2_{\rm p}\,t\,e^{-2 \kappa L},
\end{equation}
where $F$ is the cross-section of the signal beam, $t$ the
detection time interval, and $\kappa= \gamma_{13}/(v_{\rm rec} \xi)$ the
absorption coefficient introduced before. The radiative decay rate $\gamma=\gamma_1$
and the dipole matrix element $|{\mathbf d}_p|$ contained in the Rabi frequency $\Omega_p(0)$ are related 
through
\begin{equation}
\gamma=\frac{1}{4\,\pi\, \epsilon_0}\left(\frac{4}{3}\frac{|{\mathbf d}_p|^2\,\omega^3_p}{\hbar\, c^3}\right),
\end{equation}
i.~e.~according to the Einstein A-coefficient \cite{Mandel-1995}.
After a straight forward calculation we find
\begin{equation}
n_{\rm D} =F \, 
\varrho\, v_{\rm rec}\, t\, \xi\, s\, {\rm e}^{-2\,a/\xi}
\end{equation}
where $\varrho$ is the density of atoms in the EIT medium, and
\begin{equation}
a \equiv \frac{\gamma_{13} L_M}{v_{\rm rec}}
\end{equation}
characterizes the absorption due to a finite lifetime of the ground-state
coherence. Since typical values of $\gamma_{13}$ are in the kHz regime
and $v_{\rm rec}\sim 1$ cm/s, $a$ is typically large compared to unity
for $L_M\gg 10^{-3}$ cm. 
With the above expressions we find for the signal-to-noise ratio
\begin{eqnarray}
\mbox{SNR} &=& \frac{\Omega \, A}{\hbar/m} \Bigl(F \, \varrho\, v_{\rm rec}\, t\Bigr)^{1/2}\nonumber\\
&& \times \frac{\xi^{1/2} s^{1/2} (1+s)}{\xi (1+s)^3 +1}\, e^{-a/\xi}.\label{SNR}
\end{eqnarray}
The first two factors in eq.~(\ref{SNR}) are the signal-to-noise ratio of a pure
matter-wave gyroscope with interferometer area $A=R L_M$ and a flux
of atoms corresponding to a density of atoms $\varrho$ passing through an
area $F$ with recoil velocity $v_{\rm rec}$.
In conventional atomic interferometers based on cold or ultra-cold atoms the flux that contributes to the interference signal of the device is rather low. It is on the order of $10^8$ atoms/s in comparison with $10^{16}$ photons/s in a conventional fiber-optics gyroscope \cite{Kasevich-Science-2002}. However, in the case studied here, the flux can be at least two orders of magnitude higher than in an  
atom interferometer.

\begin{figure}[th]
\includegraphics[width=7.5cm]{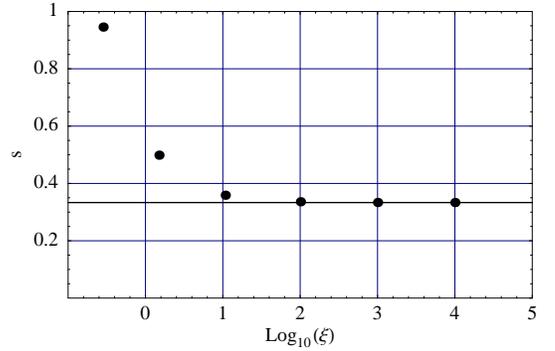}
\caption{Optimum values of $s=|\Omega_p(0)|^2/\Omega^2_c$
and $\xi=v_{\rm gr}/v_{\rm rec}-1$ for different values
of the loss parameter $a=\gamma_{13} L_M/v_{\rm rec}$ ($=0.05, 0.5, 5, 50, 500, 5000$).
For large values of $a$ the optimum values are $s_{\rm opt}=1/3$ and
$\xi_{\rm opt}=2 a$. For small values of $a$ there is only a small
deviation in the optimum parameters.}
\label{fig-4}
\end{figure}

The second factor
can be modified by optimizing the probe field strength ($s$) and the
group velocity in the medium ($\xi$). In Fig.~\ref{fig-4} we have plotted
the optimum values of $s$ and $\xi$ derived by maximizing the signal-to-noise ratio
for different values of the loss parameter $a$.

%%%%%%%%%%%%%%%%%%%%%%%%%%%%%%%%%%%%%%%%%%%%%%%%%%%%%%%%%%%%%%%%%%%%%%

%%%%%%%%%%%%%%%%%%%%%%%%%%%%%%%%%%%%%%%%%%%%%%%%%%%%%%%%%%%%%%%%%%%%%%%

One finds that in the typical parameter regime 
$a\gg 1$ the maximum SNR is attained for 
\begin{equation}
s_{\rm opt}=\frac{1}{3},\quad\textrm{and}\quad \xi_{\rm opt} = 2 a.
\end{equation}
Note that this approximation is still quite good even when $a=1$.
The optimum group velocity is given by
\begin{equation}
v_{\rm gr}^{\rm opt}=2\gamma_{13} L+v_{\rm rec}\approx 2\gamma_{13} L_M,
\end{equation}
i.e. a maximum SNR is achieved if the velocity is chosen such that
during the propagation over the entire medium length $L_M$, a fraction
of $1/\sqrt{e}$ of the initial polariton gets absorbed. Setting SNR $ = 1$
we eventually obtain the minimum detectable rotation rate
\begin{equation}
\Omega_{\rm min}= 
\frac{\hbar/m}{A} \frac{1}{\bigl(F \,\varrho\, v_{\rm rec}\, t\bigr)^{1/2}}
\, f \, \sqrt{a}
\end{equation}
where $f\approx 7.2$ is a numerical prefactor resulting from the
optimization of the term in the second line of eq.(\ref{SNR}). Apart from the 
term $\sqrt{a}$ and the unimportant
numerical prefactor $f$, the minimum detectable rotation rate
corresponds to that of a matter-wave interferometer with atoms
propagating at recoil velocity. The densities achievable in the present 
set-ups, e.g. if we consider BECs in ring trap configurations, are however 
much larger than those in  typical atomic beams.
 
 To be more precise we give an estimate for the minimum detectable rotation
rate of the slow-light gyroscope achievable with current technology.
To this end we consider two state-of-the-art circular waveguides for Bose-Einstein condensates \cite{Gupta-PRL-2005,Arnold-PRA-2006}.
Furthermore, we assume that the atomic density of the BECs is  $\rho=10^{14}$ cm$^{-3}$ 
with a cross-section (smaller circle of 
the toroidal BEC) of $F\approx 10^{-2}$ cm$^2$. In case of the work of S. Gupta et al. \cite{Gupta-PRL-2005} the diameter of the larger circle of the toroidal waveguide is $d_{\rm Gupta}\approx3$ mm and
in the case of A. S. Arnold et al. \cite{Arnold-PRA-2006} it is $d_{\rm Arnold}\approx 96$ mm. Hence, we find in the first case the minimum detectable 
rotation rate to be $\Omega_{\rm min}^{\rm Gupta}\approx 1.4 \times 10^{-9}$ s$^{-1}$ Hz$^{-1/2}$ and in the latter case
$\Omega_{\rm min}^{\rm Arnold}\approx 1.4 \times 10^{-12}$ s$^{-1}$ Hz$^{-1/2}$.
These values should be compared to the state of the art which for optical gyroscopes 
is $2 \times 10^{-10}$ rad s$^{-1}$ Hz$^{-1/2}$ \cite{Stedman-CQG-2003} and for matter-wave gyroscopes  $6 \times
10^{-10}$ rad s$^{-1}$ Hz$^{-1/2}$ \cite{Gustavson-CQG-2000}.
%
%
%%%%%%%%%%%%%%%%%%%%%%%%%%%%%%%%%%%%%%%%%%%%%%%%%%%%%%%%%%%
%
\section{Conclusion}
%
%%%%%%%%%%%%%%%%%%%%%%%%%%%%%%%%%%%%%%%%%%%%%%%%%%%%%%%%%%%
%
%
We have analyzed in detail a novel type of light-matter-wave hybrid Sagnac interferometer based on ultra-slow light in media with electromagnetic induced transparency (EIT) proposed by us in
\cite{Zimmer-PRL-2004}. In particular the influence of confining potentials was investigated and
the shot-noise limited sensitivity and the minimum detectable rotation rate determined. 
By combining features of light and
matter-wave devices the hybrid interferometer yields a minimum detectable rotation rate which is potentially 
better than the currect state of the art by up to two orders of magnitude. 
We have shown that as opposed to claims in earlier proposals for slow-light gyroscopes \cite{Leonhardt-PRA-2000}, it is not sufficient to utilize only the dispersive properties of EIT-media to achieve an enhancement of the rotation sensitivity.  It is rather necessary to employ simultaneously coherence and momentum transfer in the associated Raman transition. 
Moreover we have shown that the medium has to be prepared in a state in which it does 
not acquire any rotational phase shift. This can be achieved for example by using a superfluid BEC in a ring trap 
as EIT medium. The requirement for periodic boundary conditions reduces the potential of the hybrid interferometer idea as compared to the statements in \cite{Zimmer-PRL-2004} as it is not possible to build large area interferometers under this condition with current technology. However, the potential large flux of the proposed slow-light interferometer leads to a substantial reduction of the shot noise as compared to state-of-the-art matter-wave gyroscopes and thus leads nevertheless to a substantial sensitivity enhancement.

%%%%%%%%%%%%%%%%%%%%%%%%%%%%%%%%%%%%%%%%%%%%%%%%%%%%%%%%%
%
\section*{Acknowledgement}
%
%%%%%%%%%%%%%%%%%%%%%%%%%%%%%%%%%%%%%%%%%%%%%%%%%%%%%%%%%

F.Z. acknowledges financial support from the DFG graduate school
"Ultrakurzzeitphysik und nichtlineare Optik" at the Technical University of
Kaiserslautern.

%\bibliography{sagnac}
%\bibliography{abbrev,dipldiss,papers1910-1919,papers1920-1929,papers1930-1939,papers1940-1949,papers1950-1959,papers1960-1969,papers1970-1979,papers1980-1989,papers1990-1999,papers2000-2009,rest}
\end{document}